\documentstyle[twoside,fleqn,espcrc2,epsf]{article}

\newcommand{\AmS}{{\protect\the\textfont2
  A\kern-.1667em\lower.5ex\hbox{M}\kern-.125emS}}

\title{  Visualization of  chiral condensate at finite temperature 
         on the lattice 
         \thanks{Supported in part by FWF under Contract No. P11456} 
}

\author{Markus Feurstein, Harald Markum and Stefan Thurner \\
\vspace{3mm}
Institut f\"{u}r Kernphysik, TU Wien,
         Wiedner Hauptstra\ss e 8-10, A-1040 Vienna, Austria\\ 
}

\begin{document}

\begin{abstract}
We perform an  analysis of the topological and chiral vacuum structure 
of four-dimensional QCD on the lattice at finite temperature. 
From correlation functions we show the existence of 
local correlations between the  topological charge density and the 
quark condensate on gauge average. 
We comment on sizes of clusters of
nontrivial chiral condensate and of instantons in full QCD.  
By  analysis of individual 
gauge configurations, we demonstrate that  at the places in Euclidian 
space-time, where instantons are present, amplified production of 
quark condensate occurs. 
\end{abstract}

\maketitle


It is believed that the instantons 
as carriers of the topological charge  
might play a crucial role in understanding 
the confinement mechanism of four-dimensional QCD, if one assumes that 
they  form a so-called instanton liquid \cite{SHU88}.
Instantons have topological charge $Q$ being related to  
the zero eigenvalues of the fermionic matrix 
of a gauge field configuration 
via the Atiyah-Singer index theorem \cite{singer}.
Recently, it was demonstrated that monopole currents which constitute 
a different topological excitation of compact SU(3) gauge theory, 
appear preferably 
in the regions of non-vanishing topological charge density \cite{wir,andere}.
It has been conjectured that both instantons and monopoles are related 
to chiral symmetry breaking \cite{SHU88,MIA95,warsch96}. 
In this contribution we further 
support this idea by the following results of a direct 
investigation of the local 
correlations  
of the quark  condensate and the topological charge density on realistic 
gauge field configurations. 

For the implementation of the topological charge on a Euclidian lattice
we restrict ourselves to the so-called field theoretic definitions which
approximate the topological charge density in the continuum,
$
q(x)=\frac{g^{2}}{32\pi^{2}} \epsilon^{\mu\nu\rho\sigma}
\ \mbox{\rm Tr} \ \Big ( F_{\mu\nu}(x) F_{\rho\sigma}(x) \Big ) \ .
$
We used the plaquette and the hypercube prescription.
To get rid of  quantum fluctuations and 
renormalization constants,
we employed the Cabbibo-Marinari cooling method.
Mathematically and numerically  
the local chiral condensate $\bar \psi \psi (x)$ 
is a diagonal element of the inverse of the fermionic matrix
of the QCD action. 
We compute correlation functions between two observables ${\cal O}_1(x)$ and ${\cal O}_2(y)$ 
\begin{equation}
\label{correlations}
g(y-x)=\langle {\cal O}_1(x) {\cal O}_2(y) \rangle - 
       \langle {\cal O}_1\rangle \langle {\cal O}_2\rangle
\end{equation}
and normalize them to the smallest lattice 
separation $d_{\rm min}$, 
$ c(y-x)=g(y-x)/g(d_{\rm min})$. 
Since  topological objects with opposite sign are equally distributed,
we correlate the  
quark-antiquark density  with the square of the topological charge density.
Our simulations were performed for full SU(3) QCD on an 
$8^{3} \times 4$ lattice with
periodic boundary conditions.  Applying a standard Metropolis algorithm 
has the advantage that tunneling between sectors of different topological 
charges occurs at reasonable rates. 
Dynamical quarks in Kogut-Susskind discretization   
with $n_f=3$ flavors of degenerate  mass $m=0.1$ were taken into account using the 
pseudofermionic method. 
We performed runs  in the confinement phase at $\beta=5.2$. 
Measurements were taken on 1000 configurations separated 
by 50 sweeps.

\begin{figure}
\vspace{-1mm}
\epsfxsize=7.5cm\epsffile{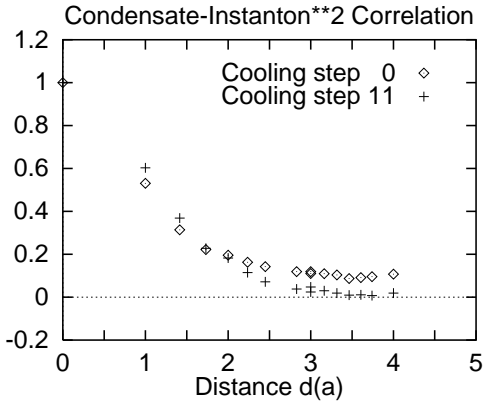 }  
\vspace{-10mm}
\caption{ 
Correlation function of the quark-antiquark density  
and the topological charge density  
for 0 and 11 cooling steps. 
The  correlations extend over two lattice spacings and indicate local 
coexistence of the quark  condensate and topological objects.
 }
\vspace{-8mm}
\label{corr}
\end{figure}
Figure~\ref{corr} shows results for the correlation functions 
of Eq.~(\ref{correlations}) with 
${\cal O}_1=\bar \psi \psi(x)$ and $ {\cal O}_2=q^2(y)$ (plaquette 
definition). 
The $\bar \psi \psi q^2$-correlation function  exhibits an extension  
of more than two lattice spacings, indicating nontrivial 
correlations. 
To gain information about the correlation lengths exponential 
fits to the tails of the correlation 
functions were performed. They show that an increasing number of cooling steps
yields shorter  correlation lengths. The corresponding 
screening masses are $\zeta=0.59$ and $\zeta=1.56$ 
in inverse lattice units for 0 and 11 cooling steps, respectively. They 
have to be interpreted as effective masses and reflect the effective gluon 
exchange and the vacuum structure of QCD.
It is assumed that the size $\rho$ 
of a t'Hooft instanton $q_\rho(x)\sim  \rho^4 \, (x^2+\rho^2)^{-4}$ 
centered around the origin enters also 
into the associated distribution of the chiral condensate 
$\bar \psi \psi_\rho (x) \sim \rho^2 \, (x^2+\rho^2)^{-3}$ 
\cite{shurev}. 
To estimate $\rho$ we fitted a convolution of the functional form 
$f(x)=\int \bar\psi\psi_\rho(t) q_\rho^2(x-t)\, dt$ 
to the data points. 
This was evaluated after 11 cooling steps where the configurations are 
reasonably dilute. Our fit yields $\rho(\bar\psi\psi q^2) 
=1.8$ in lattice spacings. 
To check consistency we extracted from the $qq$-correlation 
a value of $\rho(q q)=2.05$ which is in good agreement. 

One may choose to directly visualize densities  of the quark  condensate and
topological quantities on individual gauge fields rather than 
performing gauge averages. We persue this in the following to 
get first insight into the local interplay of topology with   
chirally nontrivial regions.  
In a series of papers we found that at the local regions
of clusters of topological charge density, which are identified with
instantons, there are monopole trajectories looping around in almost all 
cases  for  both pure SU(2) and SU(3) gauge theories \cite{wir}. 
\begin{figure*}
\vspace{-3mm}
\begin{tabular}{c|c|c}
 {\large Single Instanton} &{\large Two Antiinstantons } 
&{\large Instanton-Antiinstanton  }  \\
 & & \\
 & 0 Cooling steps & \\
\epsfxsize=4.4cm\epsffile{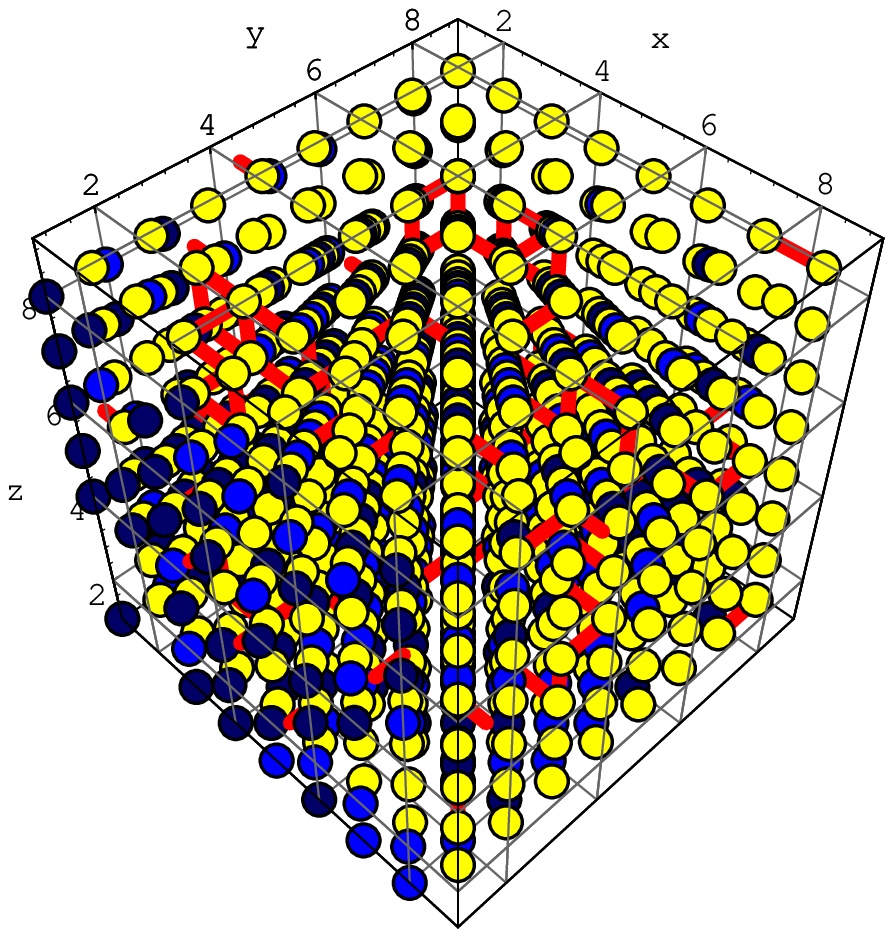}&
\epsfxsize=4.4cm\epsffile{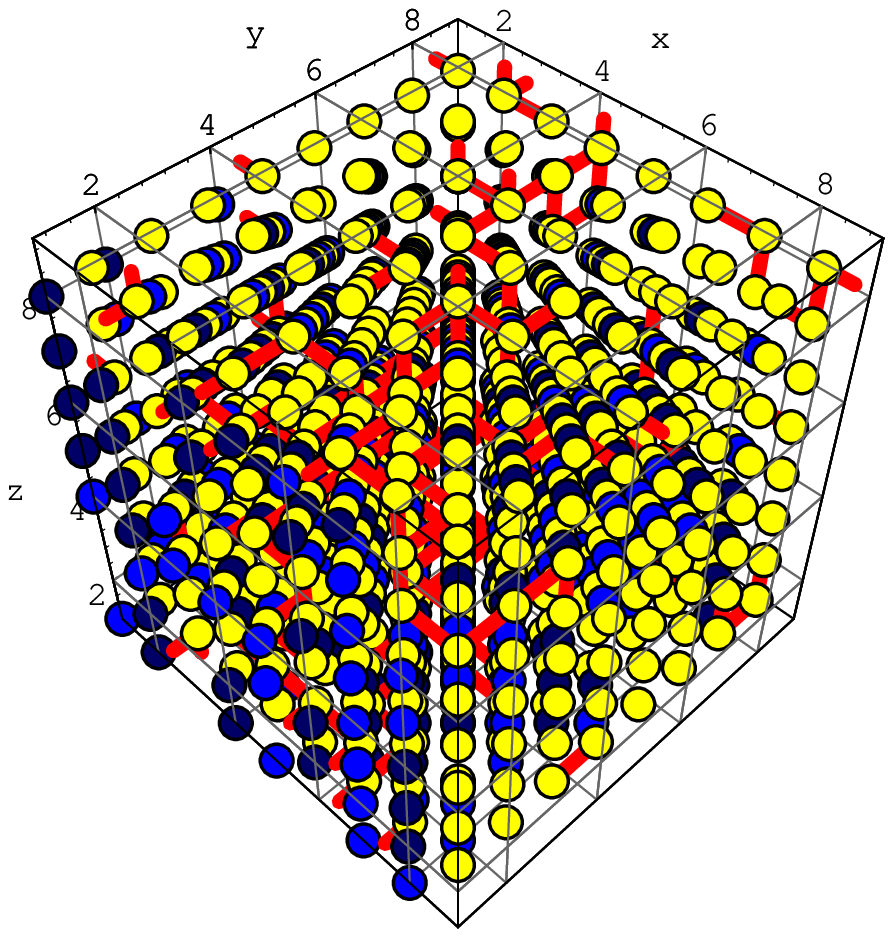}&
\epsfxsize=4.4cm\epsffile{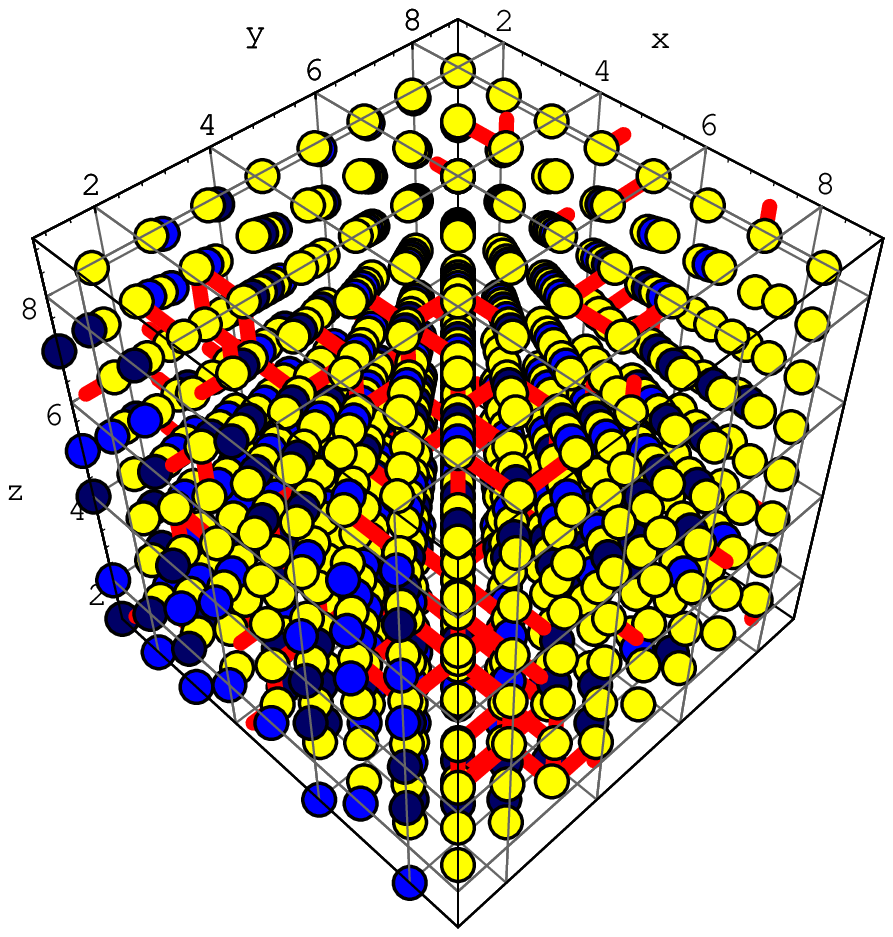}\\
 & 5 Cooling steps & \\
\epsfxsize=4.4cm\epsffile{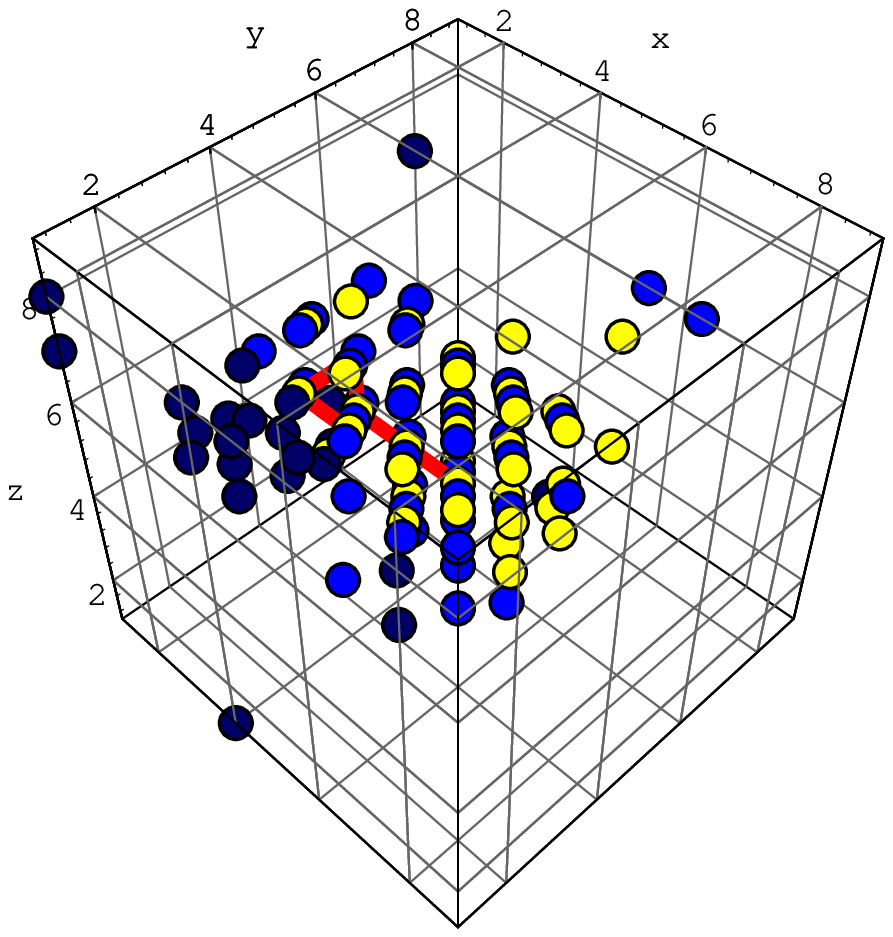}&
\epsfxsize=4.4cm\epsffile{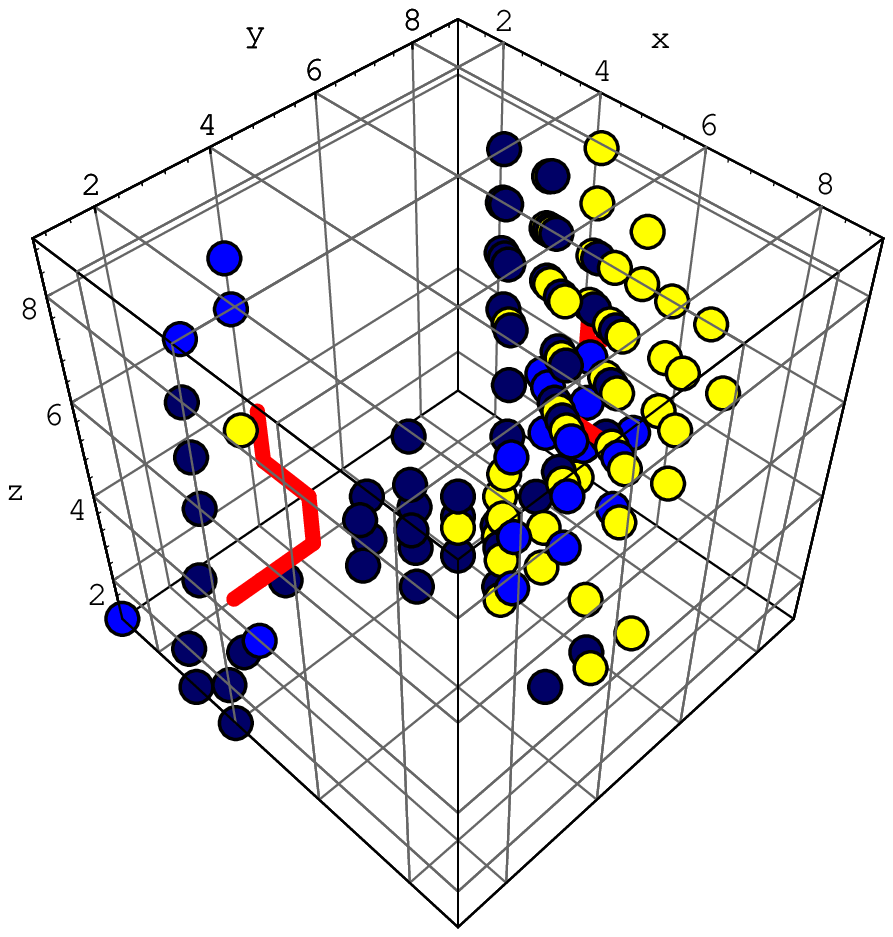}&
\epsfxsize=4.4cm\epsffile{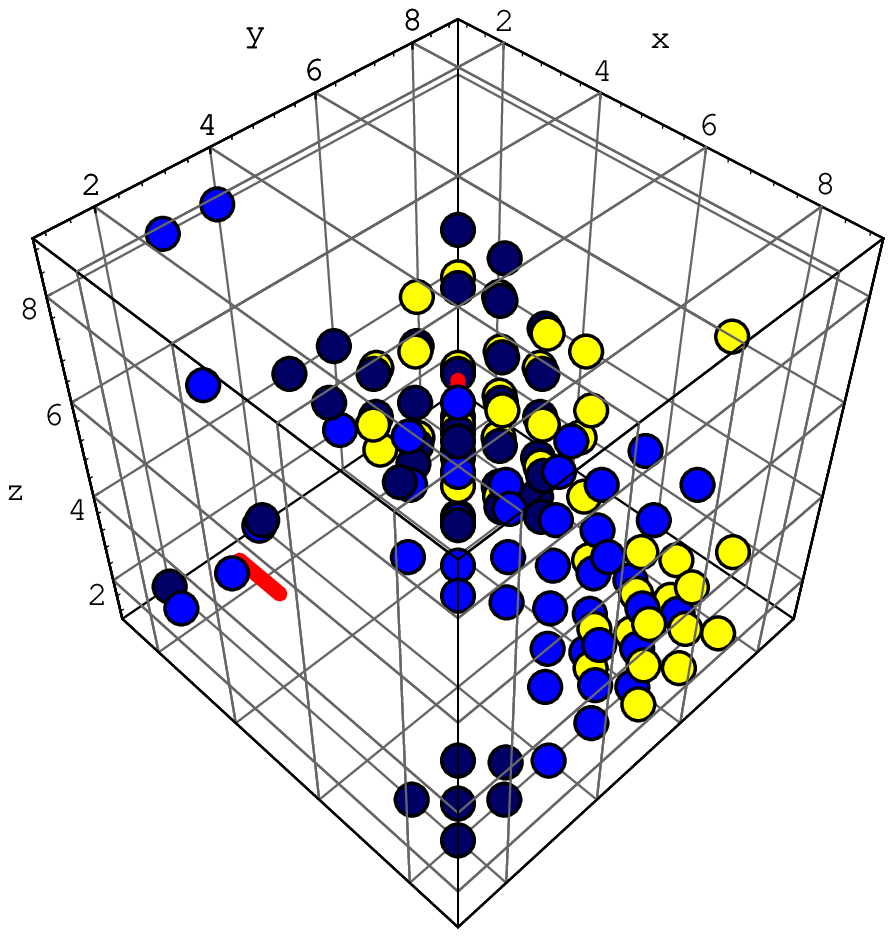}\\
 & 10 Cooling steps & \\
\epsfxsize=4.4cm\epsffile{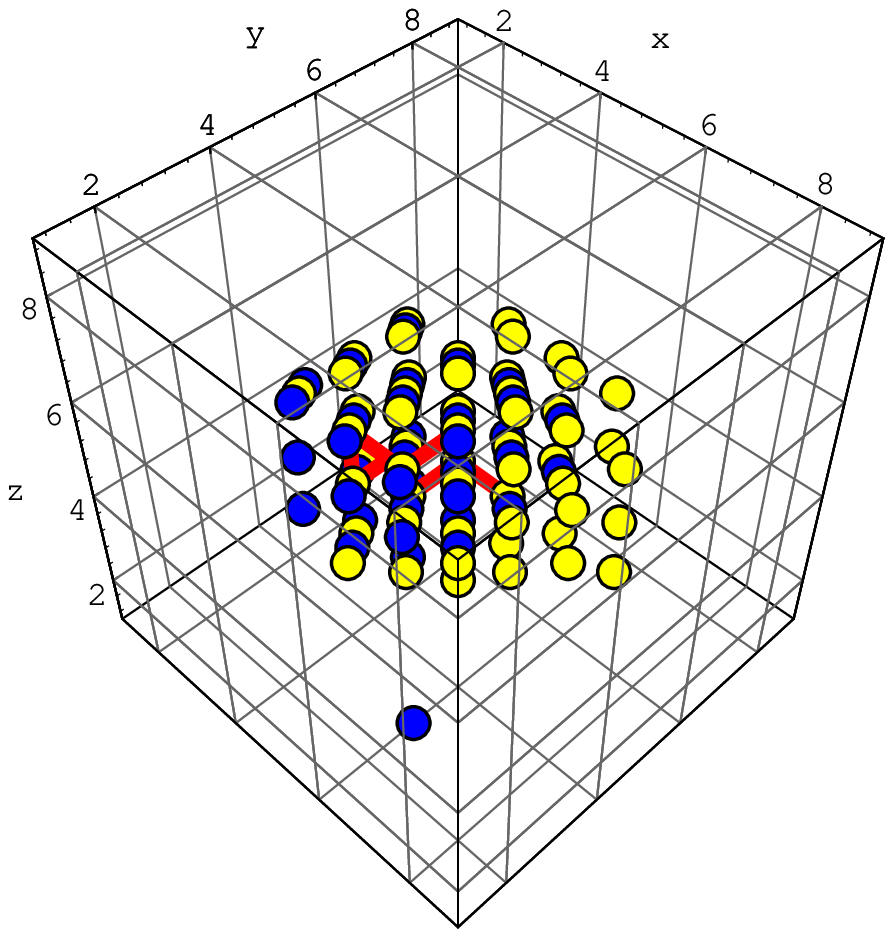}&
\epsfxsize=4.4cm\epsffile{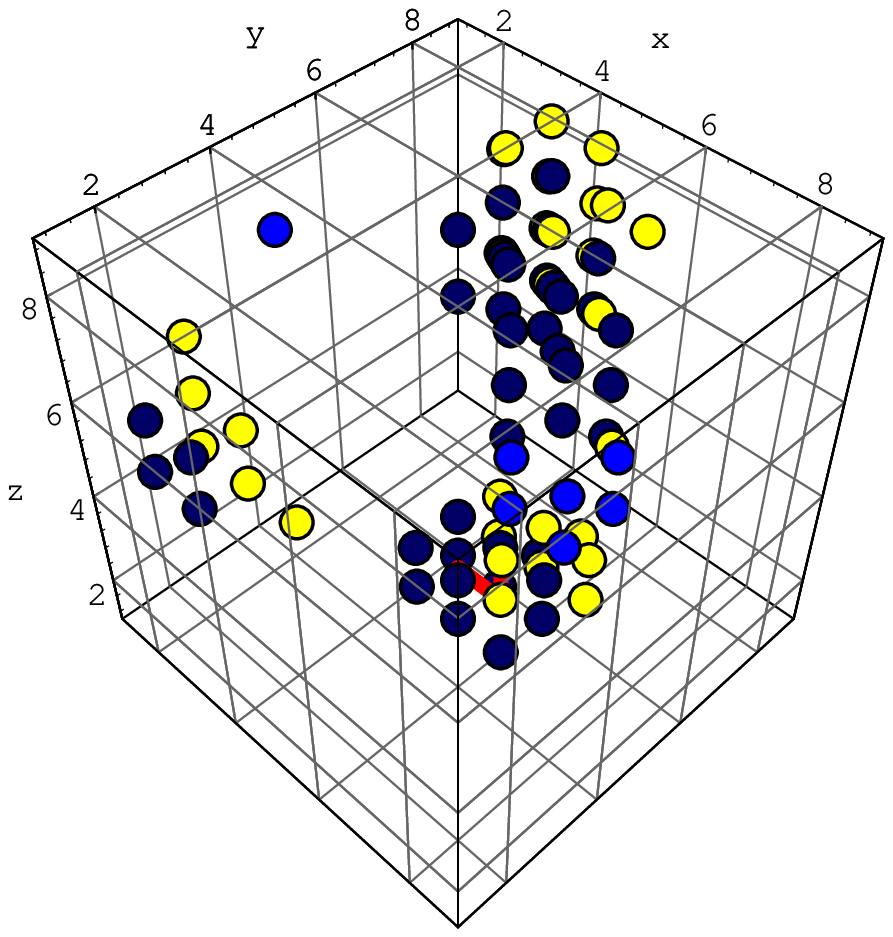}&
\epsfxsize=4.4cm\epsffile{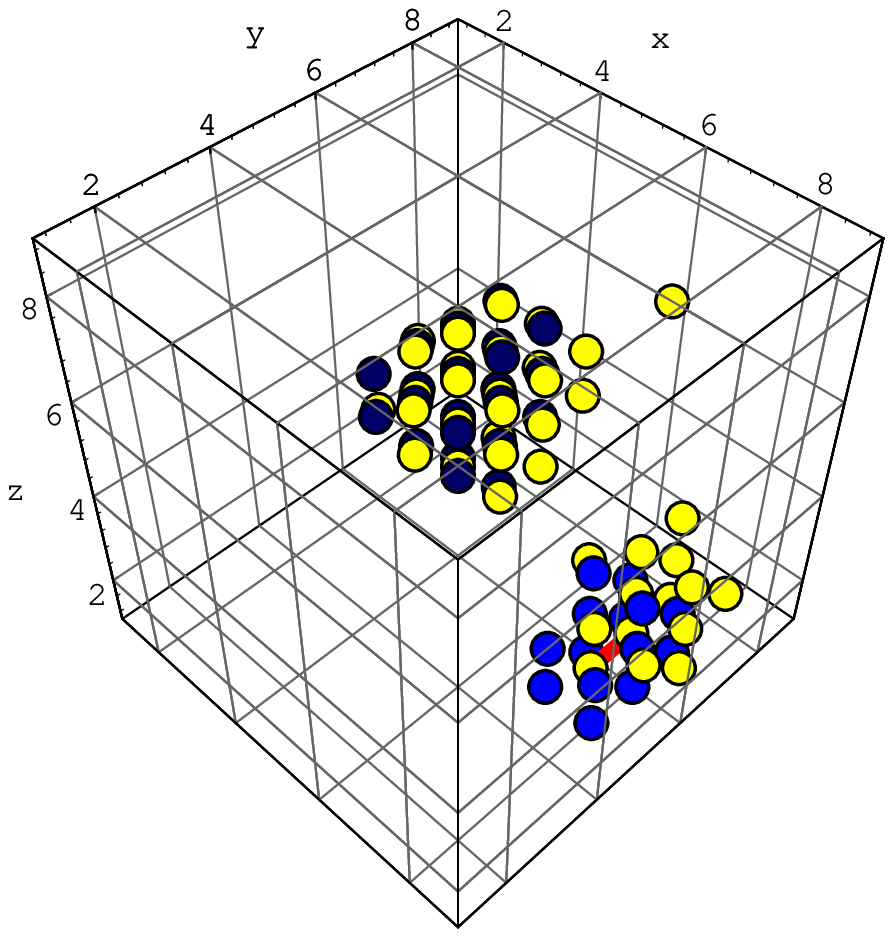}\\
 & 15 Cooling steps & \\
\epsfxsize=4.4cm\epsffile{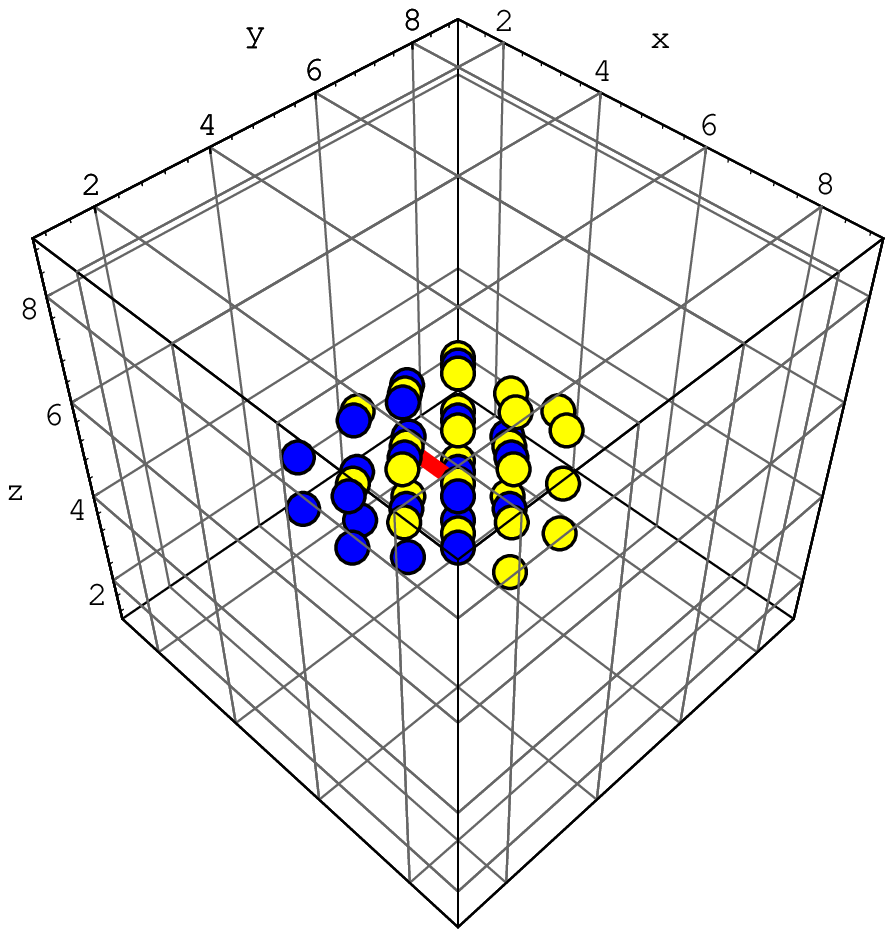}&
\epsfxsize=4.4cm\epsffile{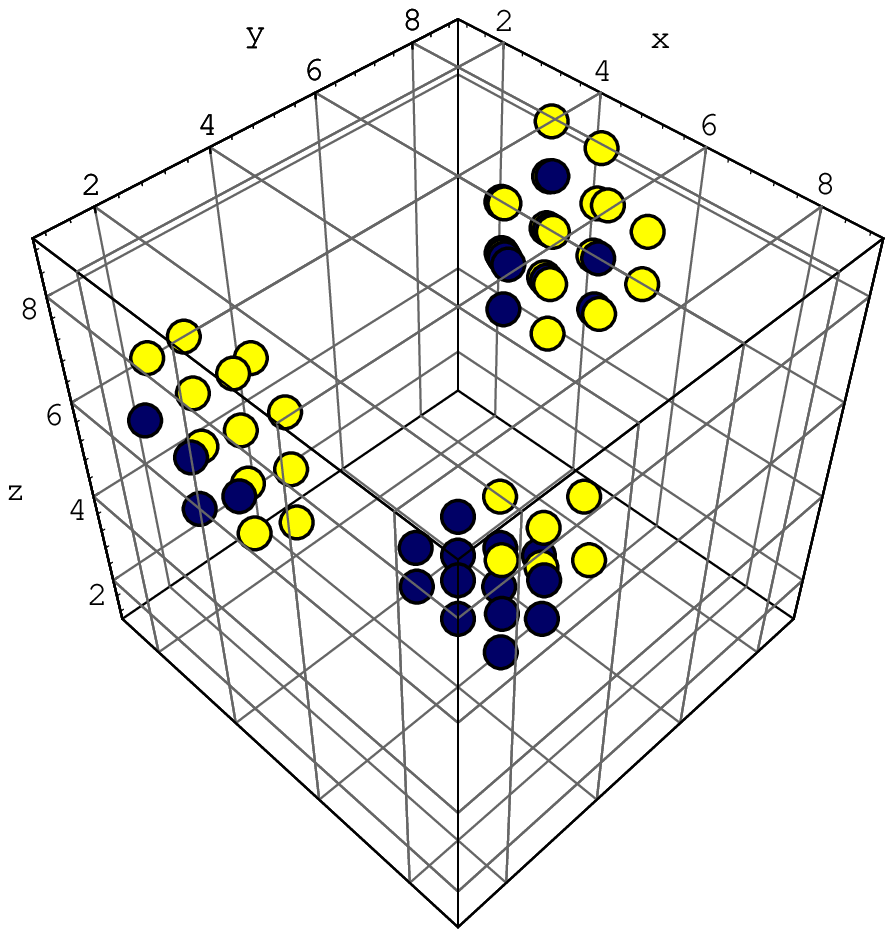}&
\epsfxsize=4.4cm\epsffile{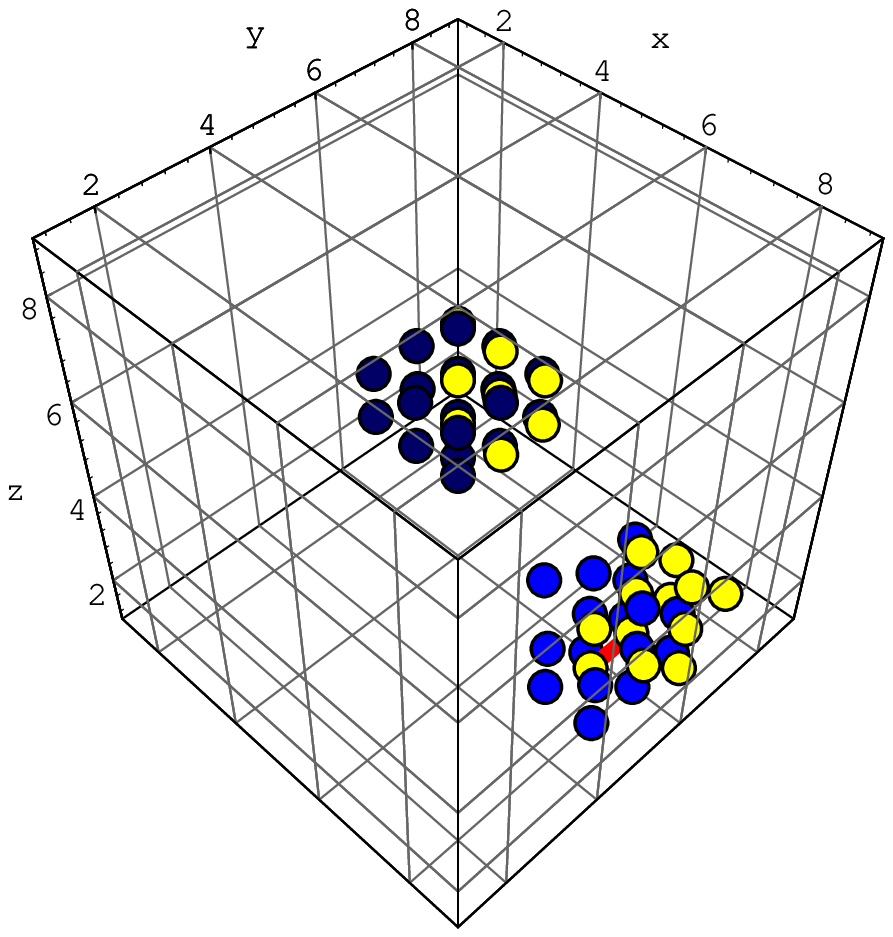}\\
\end{tabular}
\vspace{-6mm}
\caption{
Cooling history for  time slices of three  gauge field
configurations of SU(3) theory with dynamical quarks. The columns 
represent  
a single instanton, a pair of antiinstantons and an instanton-antiinstanton 
pair as the configuration gets cooled.  
The dark dots  represent the positive and negative 
topological charge density; 
the light dots  the density of the quark  condensate. 
It turns out that the quark condensate 
takes a non-vanishing value at the positions of instantons. 
  }
\label{hist}
\end{figure*}
\begin{figure*}
\vspace{-15mm}
\begin{tabular}{ccc}
\epsfxsize=5.0cm\epsffile{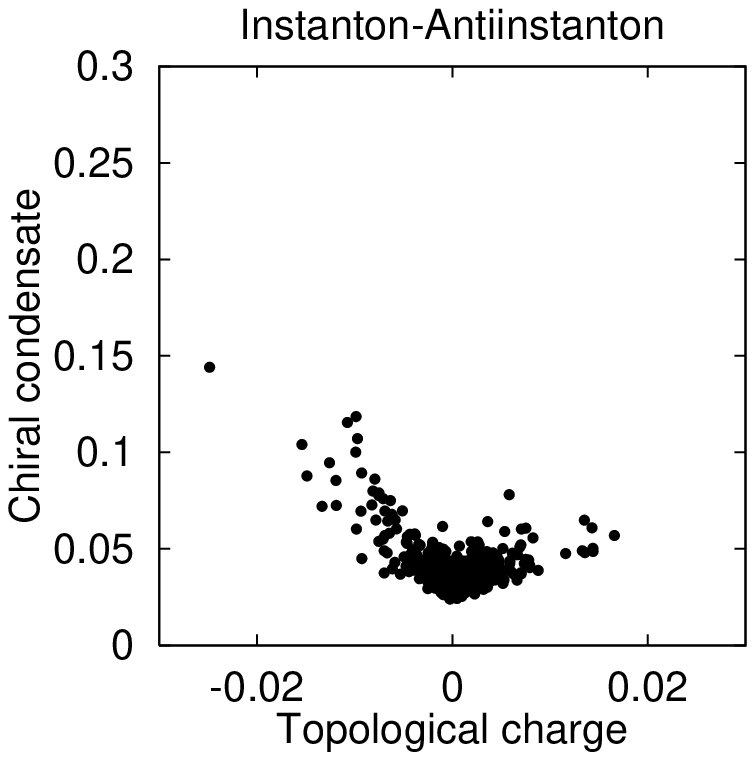} &
\epsfxsize=5.0cm\epsffile{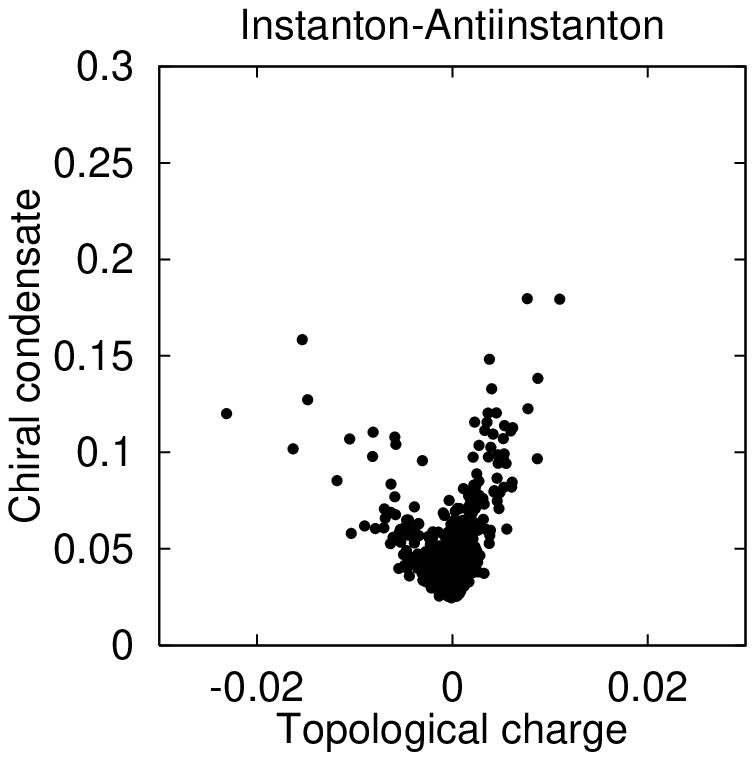} &
\epsfxsize=5.0cm\epsffile{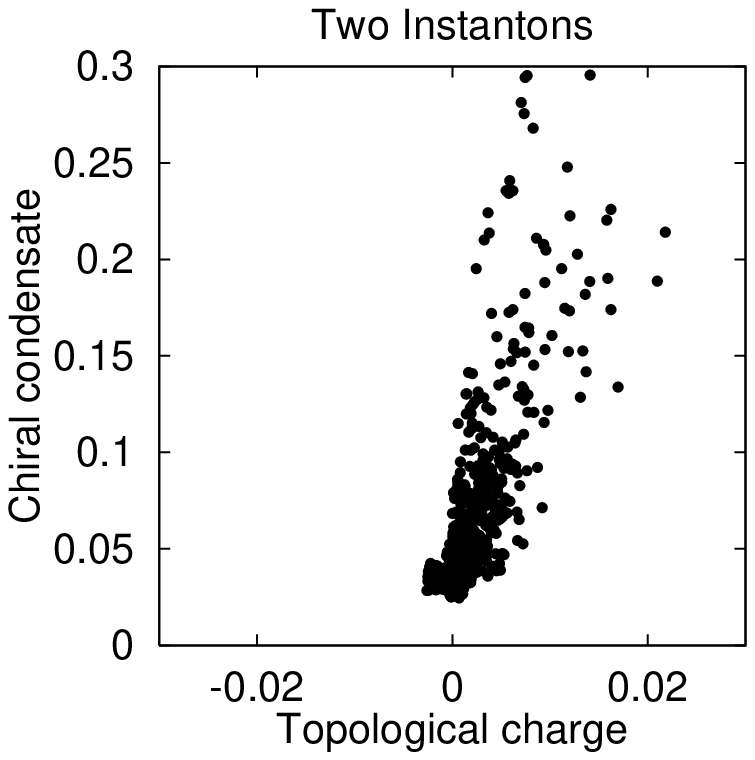} \\
\epsfxsize=5.0cm\epsffile{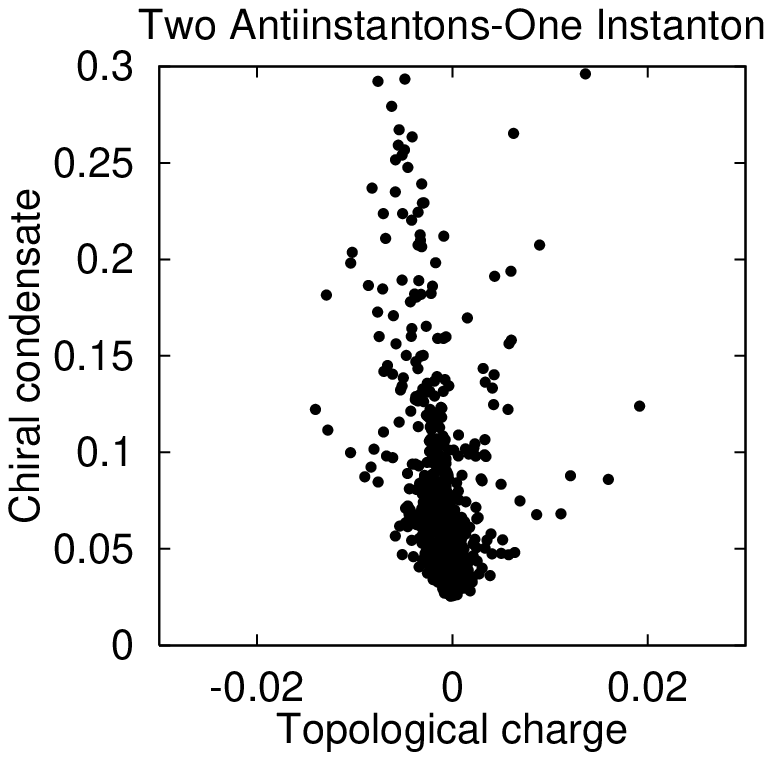} &
\epsfxsize=5.0cm\epsffile{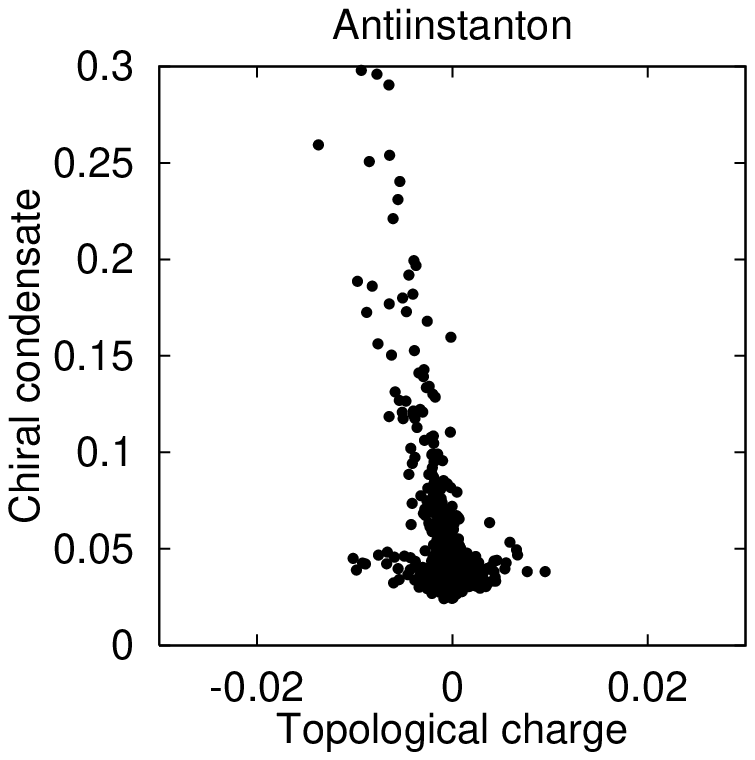} &
\epsfxsize=5.0cm\epsffile{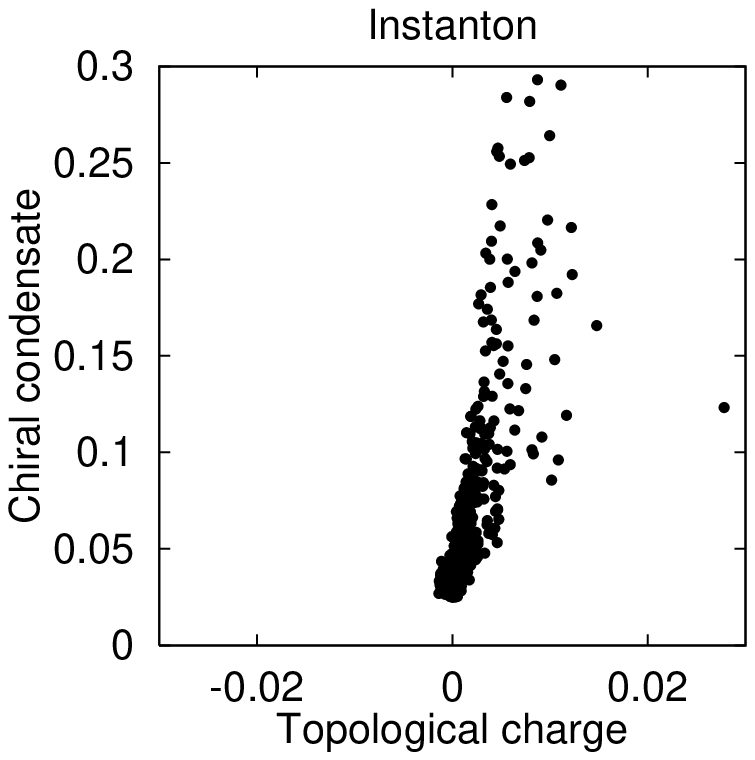} \\
\epsfxsize=5.0cm\epsffile{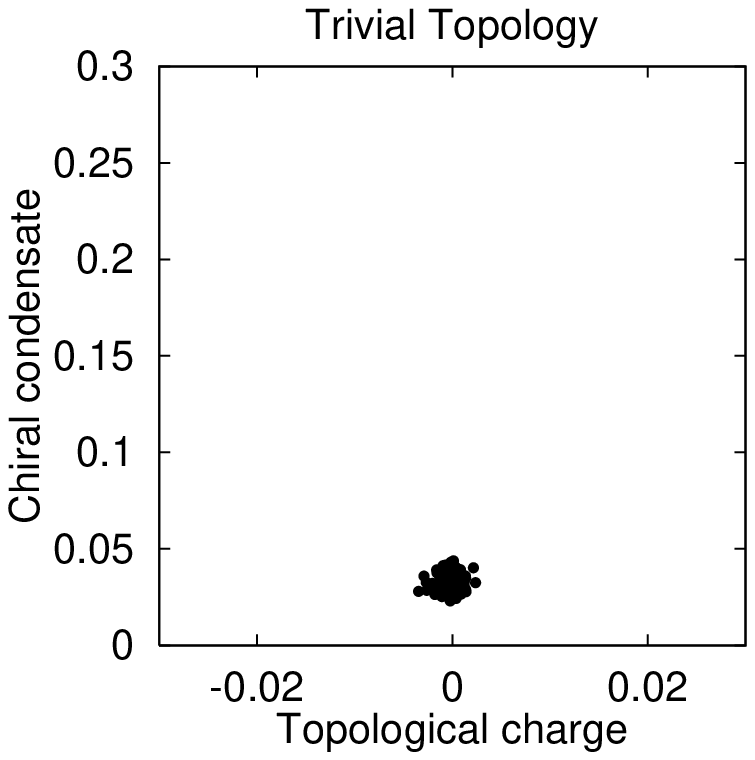} &
\epsfxsize=5.0cm\epsffile{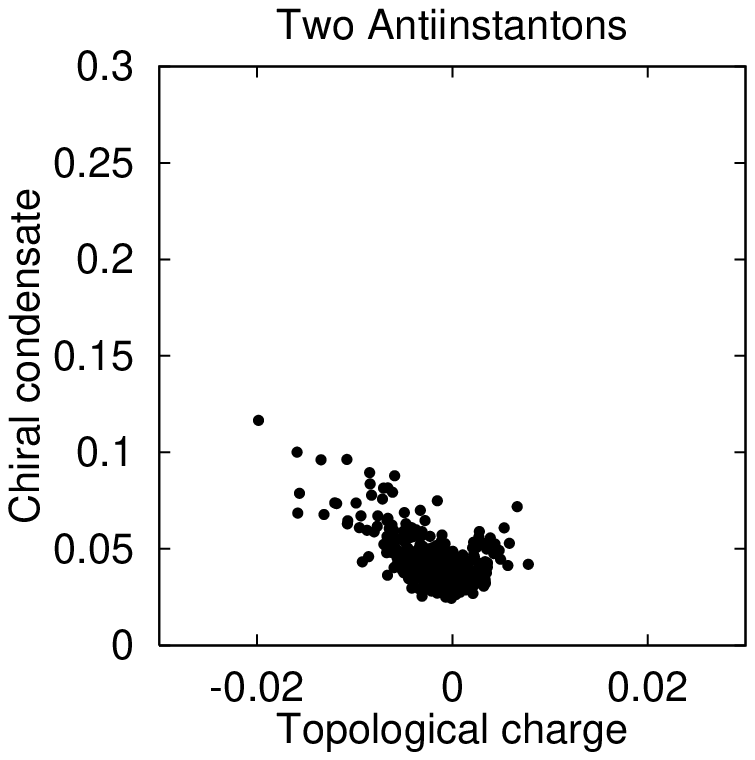} &
\epsfxsize=5.0cm\epsffile{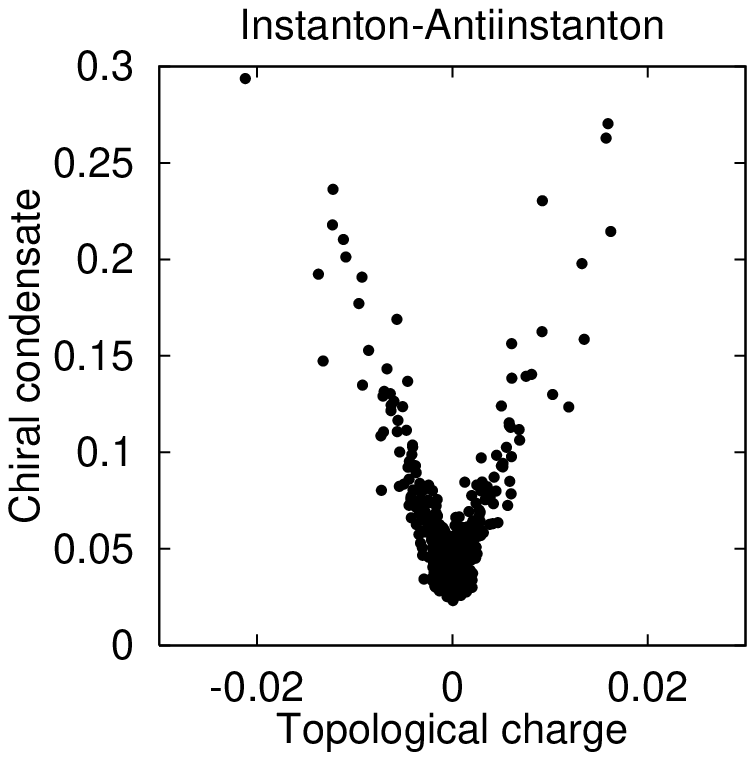} \\
\end{tabular}
\caption{Scatter plots of the local chiral condensate 
$\psi \bar \psi(x)$ and the topological charge density 
$q(x)$ for nine  configurations  
after 10 cooling steps. 
A linear relationship is suggested. 
}
\vspace{-3mm}
\label{scatt}
\end{figure*}
In Fig.~\ref{hist}    
three typical topologically nontrivial 
configurations from SU(3) theory
with dynamical quarks on the $8^{3} \times 4$ lattice 
in the confinement phase are shown for fixed time slices.
We display the topological charge density (hypercube definition) by  
dark  dots if the absolute value
$|q(x)| > 0.003$.
The quark-antiquark density  is indicated by  light dots whenever 
a threshold for $\bar \psi\psi (x) > 0.066$ is exceeded.
Monopole currents are defined in the maximum Abelian projection 
and only one type is shown  by lines. 
By analyzing dozens of gluon and quark field configurations we found the
following results.
The topological charge is hidden in  quantum fluctuations and
becomes visible by cooling of the gauge fields. For
0 cooling steps no structure can be seen in $q(x)$, $\bar \psi\psi (x)$
or the monopole currents, which does not mean the absence of correlations
between them. After about 5 cooling steps clusters of
 nonzero topological charge density and quark  condensate are resolved.
These particular configurations contain a single  
instanton, a pair of antiinstantons and an instanton-antiinstanton pair. 
For more than 10 cooling steps both topological charge and
chiral condensate begin to die out and eventually vanish.
Combining the above finding of  Fig.~1 showing that the correlation 
functions between $\bar \psi\psi (x)$ and $q^2(y)$  
are not very sensitive to cooling together with the cooling history of the 
3D images in Fig.~2, we conclude that instantons go hand in hand with clusters of 
$\bar \psi\psi (x)\neq 0$ also in the uncooled QCD vacuum 
\cite{HANDS}.

Figure~2 has demonstrated that the quark-antiquark density   attains its maximum
values at the same positions where  the extreme values of the topological
charge density are situated. 
This behavior is further substantiated in Fig.~\ref{scatt} where the
$\bar \psi\psi (x)$-values are plotted against $q(x)$ for all
points $x$. 
The nine selected configurations at 10 cooling steps have 
different topological content.  
At first sight a linear relationship between the absolute value 
of the topological charge density  and the 
virtual quark density is suggested. 
  
We comment on the behavior of the topological structure and the quark
condensate when crossing the phase transition. 
The normalized correlation functions do not change 
qualitatively \cite{wir}. The strength  of the correlations becomes,  
however, 2-3 orders of magnitude smaller. This means that the topological 
activity becomes much weaker in the deconfinement phase as expected. 
Most of the configurations have trivial net topological charge. 
This does not exclude the existence of instanton-antiinstanton pairs 
which become more difficult to be resolved at the smaller physical 
volume at $\beta=5.4$ which we also considered. 
In less than one percent
of the gauge field configurations instantons could be
identified. The configurations we scanned did not show pronounced 
pictures of instantons and quark condensate. 
The quark-antiquark density is considerably lower
and does not have  the tendency to cluster anymore, 
$\bar \psi\psi (x)$ becomes uniformly distributed over the lattice
sites. The maximum values of $\bar \psi\psi (x)$ in
configurations in the deconfinement after 10 cooling steps
are one order of magnitude smaller.

In summary, our calculations of correlation functions
between topological charge and the quark  condensate yield an
extension of about two lattice spacings. The correlations suggest that the
local chiral condensate takes a non-vanishing value predominantly in the regions  of
instantons and monopole loops.
It was well known before that the chiral condensate is related to
the topological charge and topological susceptibility. 
The visualization exhibited that the distribution of the ``chiral condensate''
concentrates around areas with enhanced topological activity 
(instantons, monopoles).
We demonstrated that exactly at those places in Euclidian space-time, 
where tunneling between the vacua occurs, amplified production of 
quark condensate takes place. 
It must be emphasized that this represents the situation on a finite lattice with 
finite quark mass without the extrapolation  to the thermodynamic and chiral limit. 
We found for full SU(3) QCD with dynamical quarks that the clusters
of non-vanishing  quark condensate have a size of about $0.4$ fm,
which corresponds to the instanton sizes
observed in the same configurations. 
Visualization of quark and gluon fields might be especially useful to decide 
if the instanton-liquid model is realized in nature and 
if instanton-antiinstanton pairs appear in the deconfined phase.
It might also help to clarify the question of the existence of a 
disoriented chiral condensate with consequences for heavy-ion 
experiments.
%

\noindent
{\bf DISCUSSIONS}
\\

\vspace{-3mm}

\noindent
{\bf Heinz J. Rothe}, Univ. Heidelberg\\
{\it 

\vspace{-7mm}

\begin{enumerate} 
\item Your results suggest that the dynamical mechanism for the 
deconfinement phase transition and chiral symmetry 
restauration  is the same. Is this also your opinion? 
 
\vspace{-2.5mm}

\item How does the cooling procedure affect the number of monopoles you see? 
\end{enumerate}
}

\vspace{-1mm}

\noindent
{\bf Harald Markum} \\
{\it

\vspace{-7mm}

\begin{enumerate}
\item  It was shown by lattice simulations that the temperature and chiral 
phase transition coincide. My opinion is that there is a common driving 
force, maybe the gluon field itself.

\vspace{-2.5mm}

\item Cooling smooths quantum fluctuations and uncovers topological 
charges. One of the caveats is that monopoles are gradually lost. 
This is a reason why one tries to develop alternative methods. 
\end{enumerate}
}

\end{document}